\documentclass[12pt,onecolumn,preprint,superscriptaddress,preprintnumbers,showpacs,tightenlines]{iopart}

\usepackage{iopams}  

\usepackage{epsfig}
\usepackage{graphicx}

\usepackage{soul}

\begin{document}

\title{Driving a mechanical resonator in to coherent states via random measurements}

\author{Ll Garcia$^1$, R W Chhajlany$^{2,3}$, Y Li$^4$ and L-A Wu$^{1,5}$}

\address{$^1$ Department of Theoretical Physics and History of Science, The Basque Country
University (EHU/UPV), P. O. Box 644, E-48080 Bilbao, Spain.}

\address{$^2$ Faculty of Physics, Adam Mickiewicz University,
PL-61-614 Pozna\'n, Poland.}

\address{$^3$ ICFO - Institut de Ci\`encies Fot\`oniques, Parc Mediterrani de la Tecnologia, E-08860
Castelldefels, Barcelona, Spain.}

\address{$^4$ Beijing Computational Science Research Center, Beijing 100084, China.}

\address{$^5$ IKERBASQUE, Basque Foundation for Science, E-48011 Bilbao, Spain.}

\eads{\mailto{lluc.garcia@ehu.es}, \mailto{ravi@amu.edu.pl}, \mailto{liyong@csrc.ac.cn}, \mailto{lianao.wu@ehu.es}}

\begin{abstract}

We propose dynamical schemes to engineer coherent states of a mechanical resonator coupled to an ancillary, superconducting flux qubit. The flux qubit, when repeatedly projected on to its ground state drives the mechanical resonator in to a coherent state in probabilistic, albeit heralded fashion. Assuming no operations on the state of the mechanical resonator during the protocol,  coherent states are successfully generated only up to a certain value of the displacement parameter. This restriction can be overcome at the cost of a one-time operation on the initial state of the mechanical resonator. We discuss the possibility of experimental realization of the presented schemes. 

\end{abstract}


\maketitle

\section{Introduction} 

The last decade has been witness to incredible progress in the fabrication
of high quality factor mechanical resonators (MR) in the nano-scale
characterized by fundamental frequencies in the 100 MHZ to 1 GHZ range.
Control of such systems in the quantum limit has emerged as a very
active field of research with implications for ultra-sensitive sensing
technologies, quantum information processing as well as fundamental
tests of quantum mechanics \cite{MR,Paraoanu}. A key required feature for
applications in the quantum limit is the ability to engineer well-defined
quantum states of the MR. The properties of a thermal MR can be manipulated
through controlled quantum mechanical coupling with electromagnetic
radiation. Indeed, this forms a powerful avenue to cool the mechanical
systems, coherently control their evolution and measure them.

Recently, various techniques have been proposed to cool the MRs to
their motional ground states, such as bang-bang cooling~\cite{bang},
single-shot state-swapping cooling via a superconductor~\cite{swap},
side-band laser cooling ~\cite{Metzger}-\cite{Licool}, electromagnetically-induced-transparency
cooling \cite{EIT-cooling}, dynamic dissipative cooling \cite{dissipative-cooling},
and through repeated projective measurements on an auxiliary flux
qubit~\cite{Lianao}. In this paper, we consider protocols for preparation
of pure coherent states, instead of ground states, of MRs from initial
general states (e.g., thermal states). Such coherent states can be
used as inputs \textit{e.g.} for schemes to generate superpositions
of spatially separated quantum states (so-called Schr\"{o}dinger cat
states) that can be used to study quantum behaviour, as well as the
transition from quantum to classical behaviour, in macroscopically
populated (phonon) states of MRs \cite{Bose}. We note here that a scheme for generating
Schr\"{o}dinger cats in a mechanical system was discussed in \cite{cat}
in the case of a MR capacitively coupled to a Cooper-pair box, which
assumes the availability of an initial coherent state without providing
for an explicit method of engineering it.



In contrast with the conventional resolved sideband cooling scheme where the MR reaches the steady state, our previous work shows that the MR cooling can be realized by repeated projective measurements on the auxiliary flux qubit \cite{Lianao}. Here, instead of ground state cooling of MR, we propose a cooling scheme to drive the MR directly into a coherent state by repeated random measurements.

\section{The driving protocol: random projective measurements}
We consider here a doubly clamped MR coupled to a gradiometer-type superconducting flux qubit~\cite{Paauw2009,flux-qubit} (the schemes presented here  formally work for coupling to any two level system). The simplest such system can be described by the effective Hamiltonian~\cite{Fei Xue}:
\begin{equation}
H=H_{q}+\hbar \omega _{m}a^{\dagger }a-\hbar g(a+a^{\dagger })\sigma _{x}.
\label{eq:1}
\end{equation}
Here $H_q$ is the free Hamiltonian of the flux qubit, the second term represents the free Hamiltonian $H_{MR}$ of the MR with fundamental frequency $\omega_m$, while the third term describes the interaction between these two systems $H_{int}$. 
In the Pauli spin-1/2 representation, the flux qubit Hamiltonian in the basis of equal superpositions of persistent currents $\left\vert\circlearrowright\right\rangle \pm \left\vert\circlearrowleft\right\rangle$ is described by a bias energy term $\sim \sigma_x$, that is set to zero by trapping a half-flux quantum in the superconducting loop~\cite{Paauw2009,Fedorov2010}, and a tunnelling term between the persistent current states, {\it i.e.} $H_q=\frac{\hbar \Delta}{2} \sigma_z$. We shall consider the weak coupling limit, {\it i.e.} $g<<\omega_m$. In this regime, the rotating wave approximation applies and the system is described by the Jaynes-Cummings model:
\begin{equation}
H_{JC}=H_{q}+ H_{MR}-\hbar g(\sigma^{\dagger}a +a^{\dagger }\sigma ^{-})
\label{JCHam}
\end{equation}
where $\sigma^{\pm}=(\sigma_{x}\pm i\sigma_{y})/2$.
In a previous paper by some of us \cite{Lianao}, based on the system
described by the above JC Hamiltonian (\ref{JCHam}), a feasible scheme
for the ground-state cooling of a MR was achieved via projective measurements
on an auxiliary flux qubit which interacts with it. 
For the protocols to generate coherent states, we shall require the MR to be displaced in space by $\alpha$.
To this end, let us assume a flux qubit-MR model
whose orginal Hamiltonian is given through a displacement transformation
$D(\alpha)\equiv\exp(\alpha a^{\dagger}-h.c.)$ to the JC Hamiltonian
(\ref{JCHam}) (up to a constant term)
\begin{equation}
H^{\rm effective}=H_{JC}-\alpha \sqrt{2m\omega_m^3}(a+a^{\dagger})-g\alpha \sigma_x.
\label{effham}
\end{equation} 
Initially, we assume that the flux qubit+MR system is prepared in the product state $|g\rangle\langle g|\otimes \rho_m$, where $|g\rangle = (\left\vert \circlearrowleft\right\rangle-\left\vert \circlearrowright \right\rangle)/\sqrt{2}$  is the ground state of the free flux-qubit Hamiltonian, while the MR is in some general state with non-zero overlap with the coherent state we want to prepare. 
The basic constituent of the protocols, we present, for driving the MR to a coherent state is a quantum-Zeno type effect: the flux-qubit+MR is allowed to undergo dynamic evolution under the Hamiltonian (\ref{effham}) which is interrupted by a series of randomly timed projected measurements on to the ground state of the flux qubit  $\mathcal{O}=\left\vert g\right\rangle\left\langle g\right\vert \otimes \mathbb{I}$. We show below that after $N$ consecutive measurements on the qubit
yielding outcomes $|g\rangle$, the state of the MR will approach 
the coherent state $|\alpha\rangle$ with fidelity increasing
to 1 as $N$ increases.
Notice that the effective evolution operator pertaining to the MR encompassing a single measurement at time $\tau$  can be disentangled to the form  
\begin{equation}
\langle g| e^{- \frac{i}{\hbar} H^{\rm effective}\tau}|g \rangle = D(\alpha) V_{g}^{(1)}(\tau) D^{\dagger}(\alpha),
\label{operator}
\end{equation}
where from (\ref{effham}) $V_{g}^{(1)}(\tau )\equiv \left\langle g\right\vert e^{-iH_{JC}\tau /\hbar}\left\vert g\right\rangle$ and furthermore $D(\alpha)=\exp(\alpha (a^{\dagger} - a))$ is the displacement operator\cite{dress}. After one measurement
\begin{eqnarray}
\mathcal{O} U \bigg(\left\vert g\right\rangle\left\langle g\right\vert \otimes \rho_m\bigg)U^{\dagger} \mathcal{O}^{\dagger}=\left\vert g\right\rangle\left\langle g\right\vert \otimes \rho_m^{\tau}(1) \\
\rho_m^{\tau}(1)=D(\alpha) V_g^{(1)}(\tau) D^\dagger(\alpha) \rho_m D(\alpha) V_g^{\dagger (1)} (\tau) D^\dagger (\alpha)
\end{eqnarray}

Thus after N consecutive measurements of the flux qubit ground state performed at times $\{\tau\}=\{\tau_1,\tau_2, \ldots \tau_N\}$, the density matrix of the MR is 
\begin{equation}
\rho_m^{\tau}(N) = \frac{D(\alpha)V_{g}^{(N)}(\{\tau\} )\,\rho^{\rm eff}_m\, V_{g}^{\dagger (N)}(\{\tau\} )D^\dagger(\alpha)}{P_{g}^{\tau}(N)},
\label{densitymr}
\end{equation}
where $V_{g}^{(N)}(\tau )= V_{g}^{(1)}(\tau_1 )V_{g}^{(1)}(\tau_2 ) \ldots V_{g}^{(1)}(\tau_N ) $ and $\rho^{\rm eff}_m = D^{\dagger}(\alpha)\rho_m D(\alpha)$. The success probability of  this process is 
\begin{equation}
P_{g}^{\tau}(N)= {\rm Tr} \; \big(D(\alpha)V_{g}^{(N)}(\{\tau\} )\,\rho^{\rm eff}_m\, V_{g}^{\dagger (N)}(\{\tau\} )D^\dagger(\alpha)\big).
\label{survival}
\end{equation}

We first analyze the properties of the above success probability. The Jaynes-Cummings evolution $\exp(-i H_{JC} \tau/\hbar)$, due to the symmetry associated with conservation of total number of excitations $a^{\dagger} a +|e\rangle\langle e| $, where $|e\rangle \sim \left\vert \circlearrowleft\right\rangle+\left\vert \circlearrowright \right\rangle$ is the excited state of the flux qubit, is easily decomposed into 2 dimensional sectors, apart from the the space of 0 excitations which is 1-dimensional \cite{Knight-book,Lianao}. The conditional evolution operator $V_g^{(1)}(\tau)$ therefore is diagonal in the phonon basis $\{|n\rangle\}$ of the MR (corresponding to $H_{MR}$):
\begin{equation}
V_g^{(1)}(\tau) = \sum_n \lambda_n |n\rangle\langle n|
\label{conditional_evolution}
\end{equation} 
with complex eigenvalues $\lambda_0= e^{i \Delta \tau/2}$ and for $n>1$,  $\lambda_{n}= e^{-i(n-1/2)\omega_{m}\tau}(\cos\Omega_{n}\tau+i\sin\Omega_{n}\tau
\cos2\theta_{n})$ 
where  $\Omega_{n}=\sqrt{(\Delta-\omega_{m})^{2}/4+g^{2}n}$ is the Rabi frequency. 
The parameter $\theta_n$ measures the ratio of the effective interaction energy scale to the deviation from resonance, according to the equation: $\tan2\theta_{n}=2g\sqrt{n}
/(\Delta-\omega_{m})$.
Note that $|\lambda_0|=1$ is  conserved due to effective decoupling of the zero phonon state during the conditional evolution, while for random times $\tau$ the coefficients 
$|\lambda_n|<1$ for higher phonon states $n>1$. 

Hence after $N$ consecutive measurements of the flux qubit ground state, the MR is driven into the state (\ref{densitymr}) given explicitly as
\begin{equation}
\rho_m^{\tau}(N)=\sum_{n,k\geqslant 0} \bar{\lambda}_n(\tau) \bar{\lambda}_k^{*}(\tau)\frac{\langle n|\rho^{\rm eff}_m |k\rangle  D(\alpha) \left\vert n\right\rangle \left\langle k\right\vert D^\dagger(\alpha)}{P_g^{(\tau)}(N)}
\label{eq:4}
\end{equation}
where $\bar{\lambda}_n(\tau) = \prod_{j=1}^{N} \lambda_n(\tau_j)$. The success probability of realizing the conditional evolution (\ref{survival}) is thus:
\begin{eqnarray}
P_g^{\tau}(N)&=&\sum_{n,k\geqslant 0} \bar{\lambda}_n(\tau) \bar{\lambda}_k^{*} \langle n|\rho^{\rm eff}_m |k\rangle  {\rm Tr}\big(D(\alpha) \left\vert n\right\rangle \left\langle k\right\vert D^\dagger(\alpha)\big) \nonumber\\
&=& \sum_{n\geqslant 0} |\bar{\lambda}_n(\tau)|^2 \rho^{{\rm eff} (n)}_m
\label{survival1}
\end{eqnarray}
where the second line comes about from the cyclic property of the trace and unitarity of the displacement operator $D(\alpha)$ and $\rho^{{\rm eff} (n)}_m=\langle n|\rho^{\rm eff}_m |n\rangle$ is the diagonal matrix element of the effective input density matrix.

On the other hand, the fidelity of the state of the MR	(\ref{eq:4}) with the coherent state $|\alpha\rangle = D(\alpha) |0\rangle$ is given by:
\begin{eqnarray}
F_g^{\tau}(N)&=& \langle 0| D^{\dagger}(\alpha) \rho_m^{\tau}(N) D(\alpha) |0\rangle\nonumber\\
&=&\sum_{n,k\geqslant 0} \bar{\lambda}_n(\tau) \bar{\lambda}_k^{*}(\tau)\frac{\langle n|\rho^{\rm eff}_m |k\rangle  \langle 0|n\rangle \langle k | 0\rangle}{P_g^{(\tau)}(N)}
\nonumber \\
&=& \rho^{{\rm eff } (0)}_m /  P_g^{\tau}(N)
\label{fidelity}
\end{eqnarray}
where in the last line we have used the fact that $|\bar{\lambda}_0 (\tau)|^2=1$. This result shows that in order to obtain a non-zero fidelity with a coherent state $\alpha$, the effective input density matrix must have a non-zero overlap with the phonon vacuum. 

We now turn to the asymptotic characteristics of our method. Due to the properties of the eigenvalues $\lambda_n(\tau)$ mentioned earlier, the products over different times $|\bar{\lambda}_n(\tau)| \rightarrow 0$ as the number of measurements $N \rightarrow \infty$ for all $n>0$ and only $|\bar{\lambda}_0(\tau)|=1$. Thus asymptotically, 
\begin{equation}
P_g^{\tau}(N) = \rho_m^{{\rm eff } (0)} = \langle \alpha|\rho_m |\alpha \rangle. 
\label{asym1}
\end{equation} 
Using this, we see from (\ref{fidelity}) that the fidelity asymptotically tends to 1, which shows that one effectively obtains the coherent state of the MR. Importantly, the results (\ref{fidelity},\ref{asym1}) also imply, that by measuring the ground state of the ancillary system, in this case the flux qubit, many times, one basically remotely implements the coherent state projective measurement on the initial state of the MR which is hard to implement directly. Indeed, (\ref{asym1}) shows that the asymptotic success probability
of the protocol is given by the probability of projecting the initial state on the target, coherent state $|\alpha\rangle$.

The  characteristics of the described process depend on the initial state $\rho_m$ of the MR. In what follows, we consider two, rather natural, possible, initializations  of the MR and compare the performance of the above described protocol.

\section{Thermal state of the MR}


Consider the initial state of the MR to be the thermal state $\rho_{m}=\exp(-\beta\hbar\omega_{m}a^{\dagger}a)/
\textrm{Tr}[\exp(-\beta\hbar\omega_{m}a^{\dagger}a)]$ $=\sum_{n}p(n)|n\rangle\langle n|$,
where the probability $p(n)=\bar{n}^{n}/(1+\bar{n}){}^{n}$ with average
phonon excitation at inverse temperature $\beta$ given by the Planck
formula $\bar{n}=1/[\exp(\beta\hbar\omega_{m})+1]$ and $\{|n\rangle\}$
is the phonon state basis of the undisplaced MR. 
This can be obained by
allowing the MR to thermalize, while the interaction with the flux
qubit ($H_{int}$) as well as the ``displacement'' term for the
MR, i.e., the second term in (\ref{effham}), is switched off.
Then the interaction as well as the displacement term are switched
on at time $t=0$ followed subsequently by measurements of the qubit
ground state.

The success probability of N consecutive measurements (\ref{survival1}) in this case becomes
\begin{equation*}
P_g^{(\tau)}(N)=\sum_{n,i\geqslant 0}\prod_{j=1}^N |\lambda_n^2(\tau_j)| p(i)\,\vert\left\langle i\right\vert D(\alpha)\left\vert n\right\rangle\vert^2
\end{equation*}
where the matrix elements of the displacement operator $D^{(i,n)}(\alpha)=\left\langle i\right\vert D(\alpha)\left\vert n\right\rangle$ reads

\begin{eqnarray}
D^{(i,n)}(\alpha) & = &\frac{1}{\sqrt{n!}}\left\langle i\right\vert D(\alpha) a^{\dagger n}\left\vert 0\right\rangle \nonumber \\ 
& = &\frac{1}{\sqrt{n!}}\left\langle i\right\vert (a^\dagger -\alpha)^n\left\vert \alpha\right\rangle \nonumber \\
& = & \frac{1}{\sqrt{n!}} \sum_{m=0}^n \left( \begin{array}{c} n \\ m \end{array} \right) \left\langle i\right\vert a^{\dagger m}\left\vert \alpha\right\rangle \alpha^{n-m}(-1)^{n-m} \nonumber \\           
& = &\exp(-|\alpha|^2)\sqrt{\frac{i!}{n!}}\,\sum_{m=0}^{{\rm min}(n,i)}\left( \begin{array}{c} n \\ m \end{array} \right) (-1)^{n-m} \frac{\alpha^{n+i-2m}}{(i-m)!}
\end{eqnarray}
\label{displacement}

In the last equality we have used the relation $(a^m\left\vert i \right\rangle)^\dagger = 0$ for $m\geqslant i+1$. We also applied this property for coherent states
\begin{equation}
\langle i-m| \alpha\rangle = \exp(- |\alpha |^2/2) \frac{\alpha^{i-m}}{\sqrt{(i-m)!}}
\end{equation}

After rearranging terms, this probability is written as
\begin{eqnarray}
P_g^{(\tau)}(N)&=&\exp(-|\alpha|^2)
\sum_n \sum_i \prod_{j=1}^N |\lambda^2_{n}(\tau_j)| p(i) \nonumber \\
&\times& \, n! \, i! \, \bigg| \sum^{{\rm min}(n,i)}_{m=0} \frac{(-1)^m \alpha^{n+i-2m}}{m!\, (n-m)!\, (i-m)!} \bigg|^2.
\label{survival2}
\end{eqnarray}

On the other hand, the fidelity (\ref{fidelity}) reads
\begin{equation}
F_g^{(\tau)}(N) = \sum_{i} p(i) p^{\alpha}(i) /P_g^{(\tau)}(N) 
\label{thermal1fid}
\end{equation}
where $p^{\alpha}(i)=\frac{\exp(-|\alpha|^2) /  \alpha^{2 i}}{i!}$ is the phonon number probability distribution in a coherent state $\alpha$.

We show below, that for this class of states, the MR can be driven into a coherent  state only for coherent parameters below some critical value $\alpha_c$ which depends on the temperature or equivalently the average phonon number of the thermal MR. Indeed, this stems from the decreasing overlap at a given temperature of the phonon number probability distribution $p(i)$ of the thermal state, which is geometric with maximum value for the vacuum, and the phonon number distribution of the coherent state $p^{\alpha}(i)$ which is Poissonian and centred at the average phonon excitation value $\alpha^2$. Thus to drive the MR into  coherent states with higher average phonon numbers will require higher temperatures. 

\section{Displaced thermal states of the MR}

For a given temperature, the restriction on the coherent state parameter can be avoided, using a scheme based on a different input state. Indeed, consider instead the thermal state corresponding to the displaced MR. Unlike in the previous scenario, we therefore consider a setup where the MR is first displaced, followed by a period of thermalization, after which the interaction is abruptly switched on followed by implementation of the driving protocol. 
In this scenario, the input state is the
thermal state of the displaced MR, $\rho_{m}^{\rm New}=$ $\exp\left[-\beta\hbar\omega_{m}\left((a^{\dagger}a-\alpha(a+a^{\dagger})\right)\right]/Z=D(\alpha)\,\rho_{m}\, D^{\dagger}(\alpha)$,
where $Z=\textrm{Tr}\left\{ \exp\left[-\beta\hbar\omega_{m}\left((a^{\dagger}a-\alpha(a+a^{\dagger})\right)\right]\right\} $ and $\rho_{m}$ is the thermal state of the undisplaced MR, as given
in the previous section.

Under this assumption, the new density operator matrix after N consecutive successful measurements, is
\begin{equation}
\rho^{(\tau)}(N) = \left\vert g \right\rangle \left\langle g \right\vert \otimes     \frac{D(\alpha)V_{g}^{N}(\tau )\,\rho_m\, V_{g}^{\dagger N}(\tau )D^\dagger(\alpha)}{P_g^{(\tau)}(N)}
\label{proposal}
\end{equation}
which is simply the displacement  of the thermal state of the (undisplaced) MR. 
 The success probability is simply
\begin{equation}
P _{g}^{(\tau )}(N)= \sum_{n\geq 0} |\bar{\lambda}_n(\tau)|^2\rho_m^{(n)}
\label{surproposal}
\end{equation}
which is independent of the coherent state parameter $\alpha$ and asymptotically (\ref{asym1}) is $\langle 0| \rho_m |0\rangle$, {\it i.e.} the occupation probability of the vacuum state of the bare MR at the given temperature. 

%

\section{Numerical results}

\begin{figure}[h!]
\center
\includegraphics[scale=0.45]{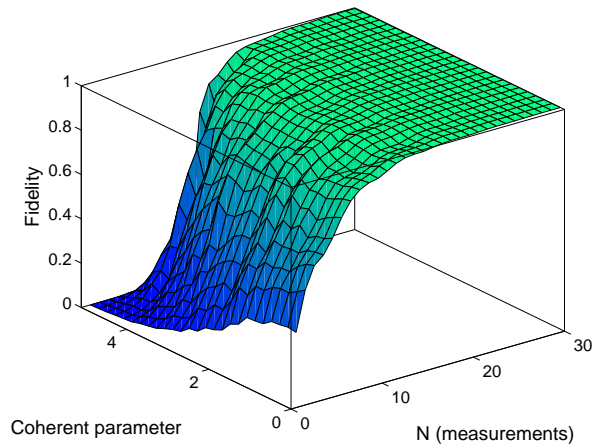} 
\caption{(Color online) Fidelity $F^{(\tau)}_g(N)$ for different values of $\alpha$ after $N$ projective measurements using $g=0.04 \,\omega_m$ and the time interval within which a single projective measurement is performed is $\tau_j =10/\omega_m$.}
 \label{plot}
\end{figure}

\begin{figure}[h!]
\center
\includegraphics[scale=0.40]{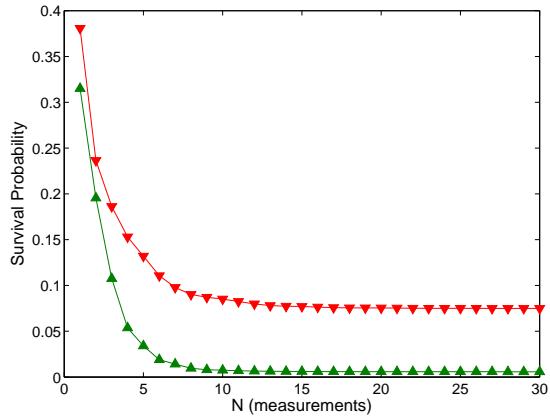} 
\caption{(Color online) The success probability $P^{(\tau)}_g(N)$ for different values of $\alpha$. Red triangles denote the case $\alpha=2$. Green triangles denote the case  $\alpha=5$. The controllable parameters are the same as in Figure~\ref{plot}.}
\label{plot2}
\end{figure}

\begin{figure}[h!]
\center
\includegraphics[scale=0.42]{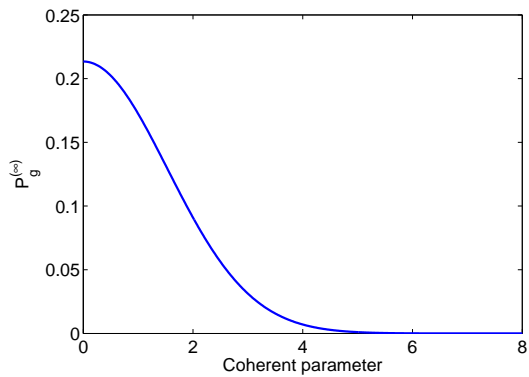} 
\caption{(Color online) Asymptotic value of the success probability depending on the displacement $\alpha$ that is sought to be obtained.}
\label{plotasym}
\end{figure}

We now study quantitative characteristics of the two presented driving protocols numerically. We consider a realistic $2\pi \times 100$ MHz nano-mechanical resonator with quality factor $Q_{m}=10^{5}$, coupled near-resonantly to a flux qubit non-resonantly with a tunnelling splitting $\Delta = 1.1\omega _{m}$. 
We first consider the scheme described in Sec. 3. The MR is taken to be in equilibrium at the ambient temperature $T=20$ mK, which corresponds to a mean phonon number of 3.69. The coupling constant is assumed $g=0.04 \omega_m$ and the measurement times $\tau_j$ are randomly selected.
Figure~\ref{plot} shows the evolution of the fidelity (\ref{fidelity}) with increasing number of projective measurements for different values of $\alpha$ up to $\alpha \leq \alpha_{\rm max}=5.1$ above which the asymptotic success probability $\leq 8 \times 10^{-4}$. 
The fidelity approaches the  asymptotic value of $1$ for all $\alpha$'s within 30 measurement steps. 
In these cases, the MR is driven close to the coherent state $\left\vert \alpha \right\rangle$ as can be seen also by considering the average phonon number  $\langle n\rangle=\delta n+|\alpha|^2$ which is the sum of thermal-like and coherent-like contributions.  The thermal-like contribution $\delta n$ can be calculated by subtracting the coherent like contribution $|\alpha|^2$ from the average phonon number $\langle n \rangle$ in the calculated final state (\ref{eq:4}). The thermal-like contribution after 30 measurements is $\delta n=2\times 10^{-2}$.
In Figure \ref{plot2}, we consider the variation of the success probability of obtaining $N$ consecutive ground state measurements of the flux qubit. For $\alpha=2$ (red curve), this probability changes very slowly after  $20$ measurements showing that the coherent state has practically been achieved. On the other hand,  as we approach the value $\alpha=5.1$  the survival probability becomes very small, {\it i.e.} we practically no longer can obtain the flux qubit in its ground state. Indeed for $\alpha=5$ (green curve) the survival probability saturates to a very small value within ten measurements.
From the results above, the asymptotic probability of success (with number of measurements $N \rightarrow \infty$) is seen to be a measure of obtaining a coherent state with fidelity $1$. It decreases to zero with increase of the coherent displacement parameter $\alpha$ as shown in Figure \ref{plotasym}. 

\begin{figure}[h!]
\center
\includegraphics[scale=0.60]{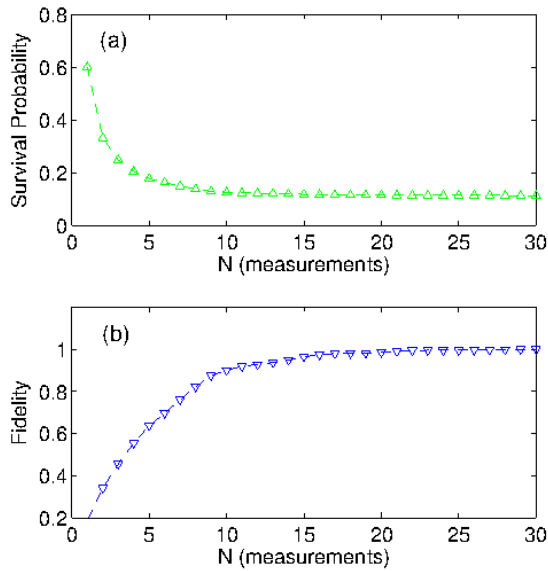} 
\caption{(Color online) (a) Survival probability after $N$ measurements. (b) Fidelity. The parameters are $T=40$ mK and $\tau=8/\omega$.}
\label{plotpropose}
\end{figure}

We now consider the scheme described in Sec. 4. 
Figure \ref{plotpropose} presents the same physical quantities as Figure~\ref{plot} and Figure~\ref{plot2} for a thermal state of the physically displaced MR. We consider a higher bath temperature $T=40$ mK with mean phonon number  7.84. After 30 measurements, the thermal-like contribution is $\delta n\simeq 10^{-3}$, thus, the average phonon number is $\langle n\rangle\simeq|\alpha|^2$. It is noticeable that in this  framework the coherent state is always reached after 30 measurements. As mentioned in the previous section, the characteristics of the protocol including the fidelity with the coherent state (see (\ref{proposal})) is independent of the displacement parameter. 

\section{Conclusions} 

We would like to remark that though the results for the protocols of Sec. 3 and Sec. 4 are obtained for the case of randomly-timed measurements, they are equally valid if one performs equal or fixed time measurements. Exact timing of measurements is however difficult to implement experimentally and indeed unfavourable. As shown in \cite{Lianao}, randomly timed measurements result in faster dynamics towards asymptotic states of the MR. 

In conclusion, we have demonstrated that quantum Zeno effect-type schemes consisting of repeated measurements on an auxiliary two level system can drive a nano mechanical resonator in to a coherent state. We propose two different ways to realize coherent state: the first one is efficient for small values of $\alpha$, while the second one, with a displaced initial state of the MR, is robust for any value of the coherent parameter.
The schemes work when the flux qubit and MR are coupled both resonantly or off-resonantly. 
We showed that the driving process is akin to performing a projective measurement on the MR without directly acting on it. 


\begin{ack}
L.A.W. and Ll. G. acknowledge the Basque Government (grant IT472-10) and the Spanish MICINN (project No. FIS2009-12773-C02-02 and No. FIS2012-36673-C03-03). R.W.C acknowledges support from the Polish National Science Centre under grant DEC-2011/03/B/ST2/01903 as well as a Mobility Plus fellowship from the Polish Ministry of Science and Higher Education. R.W.C also thanks the UPV for hospitality and funding from the AMU for a short-term visit supported by POKL Grant Nr. 04.01.01-00-133/09-00PO KL. Y. L. acknowledges support from the National Natural Science Foundation
of China (Grant No. 11174027).
\end{ack}


\end{document}